\documentclass[pre,superscriptaddress]{revtex4}

\usepackage{graphicx}
\usepackage{amssymb}
\usepackage{amsmath}

\def\3{\ss }           

\def\bomega{\mbox{\boldmath $\omega$}}

\def\bxi{\mbox{\boldmath $\xi$}}
\def\bsxi{\mbox{\boldmath \tiny $\xi$}}

\begin{document}

\title{
Equilibrium calculation of transport coefficients for a fluid-particle model} 
\author{T. Ihle}
\affiliation{Department of Physics, North Dakota State University, 
Fargo, North Dakota, 58105-5566}
\author{E. T{\"u}zel}
\affiliation{School of Physics and Astronomy, 116 Church Street SE, 
University of Minnesota, Minneapolis, MN 55455}
\affiliation{Supercomputing Institute, University of Minnesota, \\
599 Walter Library, 117 Pleasant Street S.E., \\
Minneapolis, MN 55455}
\author{D.M. Kroll}
\affiliation{Department of Physics, North Dakota State University,  
Fargo, North Dakota, 58105-5566}    
\affiliation{Supercomputing Institute, University of Minnesota, \\
599 Walter Library, 117 Pleasant Street S.E., \\
Minneapolis, MN 55455}

\begin{abstract}
A recently introduced particle-based model for fluid flow, called Stochastic 
Rotation Dynamics, can be made Galilean invariant by introducing a random 
shift of the computational grid before collisions. In this paper, it is 
shown how the Green-Kubo relations derived previously can be resummed to 
obtain exact expressions for the collisional contributions to the transport 
coefficients. It is also shown that the collisional contribution to the 
microscopic stress tensor is not symmetric, and that this leads 
to an additional viscosity. The resulting identification of the transport 
coefficients for the hydrodynamic modes is discussed in detail, and it is 
shown that this does not impose restrictions on the applicability of the model. 
The collisional contribution to the thermal conductivity, which becomes 
important for small mean free path and small average particle number per cell, 
is also derived.

\end{abstract}
\maketitle

PACS number(s): 47.11.+j, 05.40.+j, 02.70.Ns

\section{Introduction}

Particle-based simulation techniques have recently become an attractive 
alternative to more traditional methods for the coarse-grained 
modeling of a fluctuating solvent. A particularly appealing algorithm,  
introduced by Malevanets and Kapral \cite{male_99,male_00a}, and later called  
multi-particle collision dynamics \cite{lamu_01,lamu_02,ripo_04,falc_04}
or Stochastic Rotation Dynamics (SRD)  
\cite{ihle_01,ihle_03a,ihle_03b,tuze_03,tuze_04,kiku_03,pool_05,padd_04}  
provides a ``hydrodynamic heat bath,'' the details of which are not 
resolved, but which provides the correct hydrodynamic interactions 
and thermal fluctuations. The coarse-grained dynamics ignores molecular 
details, but enables simulations which span much longer time 
scales than can be addressed using molecular dynamics techniques. 
It has been used to study the behavior of polymers 
\cite{kiku_02,falc_03b,ripo_04}, colloids \cite{male_00a,falc_04,lee_04} 
(including sedimentation \cite{padd_04,hecht_05}), 
vesicles in shear flow \cite{nogu_04}, and complex fluids 
\cite{hash_00,saka_02a}. In addition to SRD's numerical advantages, its 
simplicity has made it possible to obtain analytic expressions for the 
transport coefficients which are valid for both large and small mean free 
paths, something which is very difficult to do for other mesoscale 
particle-based algorithms.  
  
In its original form \cite{male_99,male_00a}, the SRD algorithm was not 
Galilean invariant at low temperatures, where the mean free path, $\lambda$, 
is smaller than the cell size $a$. However, Galilean invariance can be 
restored by introducing a random shift \cite{ihle_01,ihle_03a}  
of the computational grid before every multi-particle interaction. 
In addition to restoring Galilean invariance, this grid shifting procedure 
accelerates momentum transfer between cells and leads to a collisional 
contribution to the transport coefficients. Two approaches 
have been used to analyze the resulting algorithm and determine the shear 
viscosity 
and thermal diffusivity. In Refs. \cite{kiku_03} and \cite{pool_05}, a 
non-equilibrium kinetic approach is used to derive the transport coefficients. 
In Refs. \cite{ihle_01,ihle_03a},    
a discrete-time projection operator technique was utilized 
to obtain Green-Kubo relations \cite{green_54,kubo_57,hansen_book} for the model's transport coefficients, 
and explicit expressions for the transport coefficients were 
derived in accompanying papers \cite{ihle_03b,tuze_03,tuze_04}. 
The two approaches are complementary and, for the most part, 
agree in their conclusions. The first is rather straightforward 
and intuitively appealing, but makes several assumptions which are not easily 
verified. The current approach justifies in detail several 
assumptions used in the non-equilibrium calculations of Refs. \cite{kiku_03} 
and \cite{pool_05}; it can also be used to analyze 
the transport coefficients of the 
longitudinal modes, namely the bulk viscosity and thermal diffusivity, which 
are hard to calculate analytically in the non-equilibrium approach 
\cite{pool_05}. Note, in particular, that the collisional contribution to 
the thermal conductivity has not yet been determined using the non-equilibrium 
methods.

In this paper, we show in detail how the time series in the Green-Kubo 
relations for the transport coefficients can be resummed in such a way as 
to eliminate all dependence on the space-fixed cell coordinates of the 
particles \cite{LETT}. This leads to a dramatic 
simplification of the sums, and enables an exact evaluation of the collisional 
contribution to the transport coefficients. It is also shown that 
there are only pure kinetic and collision contributions to the transport 
coefficients, as was implicitly assumed in the calculations of  
Refs. \cite{kiku_03,pool_05}. Explicit expressions for the collisional 
contributions to the viscosities and the thermal diffusivity are given, and 
the consequences of the fact that the collisional contribution to the 
microscopic stress tensor is {\em not} symmetric are discussed in detail.  
In particular, it is shown that this lack of symmetry leads to a slight 
change in the longitudinal viscous transport coefficient. The hydrodynamic 
modes are, however, not affected, and it does not impact on the applicability 
and stability of the method, even at low temperature.  

\section{SRD Model}

In the SRD algorithm, the fluid is modeled by particles with continuous 
spatial coordinates ${\bf r}_i(t)$ and velocities ${\bf v}_i(t)$. 
The system is coarse-grained into the cells of a regular lattice with
no restriction on the number of particles in a cell. The evolution of
the system consists of two steps: streaming and collision. In the
streaming step, the coordinate of each particle is incremented by its
displacement during the time step, $\tau$. Collisions are
modeled by a simultaneous stochastic rotation of the relative velocities
of {\em every} particle in each cell.
As discussed in Refs. \cite{ihle_01} and \cite{ihle_03a}, a random shift 
of the particle coordinates before the collision step is required to 
ensure Galilean invariance. 
All particles are shifted by the {\em same} random vector with 
components in the interval $[-a/2,a/2]$ before the collision step. Particles 
are then shifted back to their original positions after the collision. If 
we denote the cell coordinate of the shifted particle $i$ by $\bxi_i^s$, 
the algorithm is summarized in the equations  
\begin{eqnarray}
\label{eqm_1}
{\bf r}_i(t+\tau)&=&{\bf r}_i(t)+\tau\;{\bf v}_i(t) \\
\label{eqm_2}
{\bf v}_i(t+\tau)&=&{\bf u}[\bxi_i^s(t+\tau)]+\bomega[\bxi_i^s(t+\tau)]
\cdot\{{\bf v}_i(t)-{\bf u}[\bxi_i^s(t+\tau)]\},
\end{eqnarray}
where $\bomega(\bxi_i^s)$ denotes a stochastic rotation matrix, and 
${\bf u}(\bxi_i^s)\equiv\frac{1}{M}\sum_{k\in\bsxi^s}{\bf v}_k$ is the 
mean velocity of the particles in cell $\bxi^s$. All particles in the 
cell are subject to the same rotation, but the rotations in different cells 
are statistically independent. There is a great deal of freedom in how the 
rotation step is implemented, and any stochastic rotation matrix consistent 
with detailed balance can be used. 
In two dimensions, the stochastic rotation matrix, $\bomega$, is typically 
taken to be a rotation by an angle $\pm\alpha$, with probability $1/2$ 
(see Refs. \cite{ihle_01,ihle_03a,ihle_03b}). In three dimensions, two 
collision rules have been considered. In the first, (Model A in Ref.~\cite{tuze_03}),  
one performs rotations by an angle $\alpha$ about a randomly chosen direction, where all 
orientations of the random axis occur with equal probability. In the second, 
(Model B in Ref.~\cite{tuze_03}), rotations are performed about one of three    
orthogonal rotation axes, i.e. $x$-, $y$- and $z$-axes of a 
cartesian coordinate system.   
At each collision step one of these three axes is chosen at random, and a 
rotation by an angle $\pm\alpha$ is then performed, where the sign is 
chosen at random. 
	
\section{Transport Coefficients} 

Because of the cell structure introduced to define coarse-grained collisions,  
angular momentum is not conserved in a collision \cite{pol_fl,aris_89}. As a consequence, the 
macroscopic viscous stress tensor is not, in general, a symmetric function 
of the derivatives $V_{\alpha\beta}\equiv\partial_\alpha v_\beta$. 
Its general form can be determined as follows. Both the macroscopic viscous 
stress tensor, $\hat \sigma_{\alpha\beta}$, and the velocity gradient tensor,
$V_{\alpha\beta}$ which appear in the Navier-Stokes equation are rank 
two tensors. If the velocity gradients are small, we can assume---as is 
generally done---that the momentum transfer due to viscosity depends only 
on the first derivatives of the velocity, so that  
\begin{equation}  
\hat \sigma_{\alpha\beta}=C_{\alpha\beta\gamma\rho}\, V_{\gamma \rho}. 
\end{equation} 
Symmetry arguments can be used to reduce the number 
of independent elements in the compliance tensor $C_{\alpha\beta\gamma\rho}$. 
Because of the simple cubic grid structure used in the algorithm, we have 
at least cubic symmetry. In this case, it can be shown \cite{aris_89} that 
the most general form for $C_{\alpha\beta\gamma\rho}$ is
   \begin{equation} 
C_{\alpha\beta\gamma\rho}/\rho=a\ \delta_{\alpha\beta}\delta_{\gamma\rho} 
+ b\ \delta_{\alpha\gamma}\delta_{\beta\rho} + c\   
\delta_{\alpha\rho}\delta_{\beta\gamma} + \epsilon 
\Gamma_{\alpha\beta\gamma\rho},     
\end{equation} 
\noindent 
where $\tensor{\Gamma}$ is the rank four unit tensor \cite{aris_89}, $a$, $b$, $c$ and $\epsilon$ are viscosity coefficients, and $\rho$ is the mass density of the fluid. 
It follows that 
\begin{eqnarray} 
{\hat \sigma_{\alpha\beta}}/\rho &=&a\ \delta_{\alpha\beta} \partial_\lambda 
v_\lambda + b\ \partial_\alpha v_\beta + c\ \partial_\beta v_\alpha + 
\epsilon \Gamma_{\alpha\beta\gamma\rho} \partial_\gamma v_\rho \nonumber\\    
&=&\nu_1(\partial_\alpha v_\beta + \partial_\beta v_\alpha - \frac{2}{d} 
\delta_{\alpha\beta} \partial_\lambda v_\lambda)+ 
\nu_2(\partial_\beta v_\alpha - 
\partial_\alpha v_\beta) +  
\gamma \delta_{\alpha\beta} \partial_\lambda v_\lambda+  
\epsilon \Gamma_{\alpha\beta\gamma\rho} \partial_\gamma v_\rho \;\;, \label{stress}
\end{eqnarray}
\noindent with kinematic shear viscosities $\nu_1\equiv(b+c)/2$, 
$\nu_2\equiv(c-b)/2$, and 
bulk viscosity $\gamma=a+(b+c)/d$, where $d$ is the spatial dimension.
$\nu_2$ is the viscous transport coefficient associated with the non-symmetric 
part of the stress tensor, and $\epsilon$ is a viscosity coefficient related to a possible lack of full rotational symmetry; both 
$\nu_2$ and $\epsilon$ 
are usually zero in simple fluids. 
If we define a new tensor from the diagonal elements of $V_{\alpha\beta}$,
namely $R_{\alpha \beta}\equiv \Gamma_{\alpha \beta \gamma \rho} 
V_{\gamma \rho}$,
the resulting form of the momentum equation for a fluid reads
\begin{eqnarray} 
\rho\left[\frac{\partial{\bf v}}{\partial t}+({\bf v}\cdot\nabla){\bf v}
\right] &=& -\nabla p + \nabla \cdot \tensor{\hat \sigma} \\
&=& -\nabla p + \rho (\nu_1 + \nu_2) \Delta{\bf v}
+ \rho \left( (1-\frac{2}{d}) \nu_1 -  \nu_2 + \gamma \right) 
\nabla (\nabla\cdot{\bf v}) + \epsilon \nabla \cdot \tensor{R}\;\;,
\label{SHEAR1}
\end{eqnarray} 
where $p$ is the pressure.
It can be seen from Eq. (\ref{SHEAR1}) that effective shear viscosity is 
$\nu=\nu_1+\nu_2$. In momentum space, the resulting linearized Navier-Stokes 
equation can be written as 
\begin{equation} 
\partial_t v_\alpha({\bf k}) = -\frac{\partial_\alpha p}{\rho} +  
\Lambda_{\alpha\beta}(\nu_1,\nu_2,\gamma,\epsilon;\hat{\bf k}) 
v_\beta({\bf k}), 
\end{equation}  
where 
\begin{equation}
\label{ST} 
\Lambda_{\alpha\beta}(\nu_1,\nu_2,\gamma,\epsilon;{\bf \hat k}) \equiv 
\nu_1\left(\delta_{\alpha\beta}+\frac{d-2}{d}\hat k_\alpha \hat k_\beta\right) + 
\nu_2\left(\delta_{\alpha\beta} - \hat k_\alpha \hat k_\beta \right) 
+ \gamma \hat k_\alpha \hat k_\beta 
+ \epsilon\ \hat k_\gamma \hat k_\rho \Gamma_{\alpha\beta\gamma\rho}.  
\end{equation} 
$\Lambda_{\alpha\beta}(\nu_1,\nu_2,\gamma,\epsilon;{\bf \hat k})$ is the 
matrix of viscous transport coefficients. In a simple liquid, 
$\epsilon=0$ (because of invariance with respect to infinitesimal rotations),  
$\nu=\nu_1$, and $\nu_2=0$ (because the stress tensor is symmetric in 
$\partial_\alpha v_\beta$). In this case, Eq. (\ref{ST}) reduces to the well 
known form \cite{ihle_03a} 
\begin{equation} 
\Lambda_{\alpha\beta}(\nu_1,\nu_2,\gamma,\epsilon;{\bf \hat k}) 
=\nu\left(\delta_{\alpha\beta}+\frac{d-2}{d}
\hat k_\alpha \hat k_\beta\right) + \gamma \hat k_\alpha \hat k_\beta . 
\end{equation} 

As shown in Ref. \cite{ihle_03a}, the discrete Green-Kubo (GK) relation 
\begin{equation} 
\label{VTCGK} 
\Lambda_{\alpha\beta}(\nu_1,\nu_2,\gamma,\epsilon;\hat{\bf k}) \equiv 
\frac{\tau}{Nk_BT}\left.\sum_{n=0}^\infty\right.' 
\langle\hat k_\lambda \sigma_{\alpha\lambda}(0)\vert 
\hat k_{\lambda'} \sigma_{\beta\lambda'}(n\tau)\rangle    
\end{equation} 
for the SRD model 
can be used to expresses the matrix of viscous transport coefficients in 
terms of a sum of time correlation functions of the reduced fluxes  
$I_{1+\alpha}(\hat{\bf k},t) $
\begin{equation} 
\label{VIS2}   
\hat k_\lambda\sigma_{\alpha\lambda}(t) \equiv 
I_{1+\alpha}(\hat{\bf k},t) = \frac{1}{\tau}\sum_j\left(-[v_{j\alpha}(t) 
{\bf \hat k}\cdot\Delta\bxi_j(t) + \Delta v_{j\alpha}(t)
{\bf \hat k}\cdot\Delta\bxi_j^s(t)] + 
\frac{\tau \hat k_\alpha}{d}v_j^2(t)\right) 
\end{equation} 
for $\alpha=1,\dots,d$,   
with $\Delta\bxi_j\left(n\tau\right) = 
\bxi_j\left(\left[n+1\right]\tau\right) - 
\bxi_j\left(n\tau\right)$, $\Delta\bxi^s_j\left(n\tau\right) = 
\bxi_j\left(\left[n+1\right]\tau\right) - 
\bxi^s_j\left(\left[n+1\right]\tau \right)$, 
and 
$\Delta v_{xj}(n\tau)=v_{xj}\left(\left[n+1\right]\tau\right)-v_{xj}(n\tau)$. 
$\bxi_j(n\tau)$ is the cell coordinate of particle $j$ at time $n\tau$, 
while $\bxi_j^s$ is it's cell coordinate in the (stochastically) shifted
frame. 
The prime on the sum indicates that the $t=0$ term has the relative weight 
$1/2$. The sum in Eq. (\ref{VIS2}) runs over all $N$ particles of the system.

The corresponding expression for the thermal diffusivity is \cite{ihle_03a}  
\begin{equation}\label{DT} 
D_T = \frac{\tau}{c_pNk_BT^2} 
\left.\sum_{n=0}^\infty \right.'
\langle \hat k_\lambda \chi_\lambda(0) \vert 
\hat k_{\lambda'}\chi_{\lambda'}(n \tau)\rangle , 
\end{equation}  
with the energy flux 
\begin{equation}
\label{DT1} 
\chi_\lambda(n\tau)=
\frac{1}{\tau}\sum_j \left[ \left(c_v T -v^2_{j}(n \tau)/2 \right) 
\Delta\xi_{j\lambda}(n\tau) -\frac{1}{2}\Delta v_j^2(n \tau) 
\Delta\xi^s_{j\lambda}(n\tau) + \tau k_B T v_{j\lambda}(n \tau)\right],   
\end{equation}
where $c_v=d k_B/2$ is the specific heat per particle at constant volume of 
an ideal gas and 
$\Delta v_j^2(n\tau)=v_j^2\left(\left[n+1\right]\tau\right) -v_j^2(n \tau)$. 
The thermal conductivity, $\kappa$, is related to $D_T$ by 
$\kappa=\rho c_p D_T$. Here and in the following we have set the particle 
mass equal to one.  

\section{Resummed Green-Kubo relations}

The straightforward evaluation of the GK relations 
\cite{green_54,kubo_57,hansen_book} presented in Ref. \cite{ihle_03b} leads 
to three contributions to the transport coefficients, which were called the 
kinetic, rotational (or collisional), and mixed terms. The term ``rotational'' 
and the superscript ``rot'' refer to contributions from the 
collisions---stochastic rotations of the relative particle 
velocities---in the SRD-model. For large
mean free path, $\lambda\gg a$, $\lambda=\tau\sqrt{k_B T}$, the assumption of molecular chaos is valid,
and the kinetic contribution could be determined explicitly. For mean
free paths $\lambda$ smaller than the cell size $a$, however, there were finite cell
size corrections, and it was not possible to sum these contributions
in a controlled fashion. The origin of the problem was the explicit
appearance of $\Delta \xi$ in the stress correlation functions.

In fact, the appearance of $\Delta \xi$ is troubling, since one would
not expect this if the cell shifting procedure really does
restore Galilean invariance. The key to resolving this dilemma is to
realize that a proper resummation of the GK relations removes this dependence.
Consider first the time series 
$\left.\sum_{n=0}^\infty \right.' \hat k_\lambda 
\sigma_{\alpha\lambda}(n \tau)$. By canceling 
$\xi$-dependent terms in successive contributions to this series, it can be 
shown that 
\begin{equation} 
\left.\sum_{n=0}^\infty \right.' \hat k_\lambda\sigma_{\alpha\lambda}(n \tau) 
= \frac{1}{2}\hat k_\lambda [A_{\alpha\lambda}(0)+A_{\alpha\lambda}(\tau)] 
+ \left.\sum_{n=0}^\infty \right.'
\hat k_\lambda \bar\sigma_{\alpha\lambda}(n\tau) , 
\end{equation} 
where $\bar\sigma_{\alpha\lambda}(n \tau)\equiv
\bar\sigma_{\alpha\lambda}^{kin}(n \tau) +
\bar\sigma_{\alpha\lambda}^{rot}(n \tau)$ with    
\begin{equation}
\label{VIS2N}
\bar\sigma^{kin}_{\alpha\lambda}(n\tau)=-\sum_j [v_{j\alpha}(n\tau)
v_{j\lambda}(n\tau) - \delta_{\alpha\lambda}v_j^2(t)/d],  \ \ \ \ \  
\bar\sigma^{rot}_{\alpha\lambda}(n\tau)=
-\frac{1}{\tau}\sum_j v_{j\alpha}(n\tau)B_{j\lambda}(n\tau), 
\end{equation} 
\begin{equation}\label{A_def} 
A_{\alpha\lambda}(\tau) \equiv \frac{1}{\tau}\sum_j v_{j\alpha}(\tau)
\Delta\xi^s_{j\lambda}(0),  
\end{equation} 
and 
\begin{equation}
\label{B_def}
B_{j\lambda}(n \tau)\equiv \xi_{j\lambda}^s\left(\left[n+1\right]\tau\right)
                    -\xi_{j\lambda}^s(n\tau)-\tau v_{j\lambda}(n\tau) = 
               \Delta\xi_{j\lambda}(n\tau)-\Delta\xi_{j\lambda}^s(n\tau) + 
               \Delta\xi_{j\lambda}^s([n-1]\tau)-\tau v_{j\lambda}(n\tau).  
\end{equation}
$B_{j\lambda}$ is a new stochastic variable which has very simple temporal 
correlations describing the geometrical properties of the underlying lattice.  

Similarly, it can be shown that 
\begin{equation} 
\sigma_{\alpha\lambda}(0) = A_{\alpha\lambda}(0)-A_{\alpha\lambda}(\tau)
+\bar\sigma_{\alpha\lambda}(0) .   
\end{equation}  
Using these results, the Green-Kubo relation (\ref{VTCGK}) for the viscous 
transport coefficients can be written as the sum of two terms, 
\begin{eqnarray}
\label{SGT1}
\frac{\tau \hat k_\lambda\hat k_{\lambda'}}{N \,k_B T}\left\langle 
\frac{1}{2}[A_{\alpha\lambda}(0) A_{\beta\lambda'}(0) - A_{\alpha\lambda}(\tau)A_{\beta\lambda'}(\tau)]
+\frac{1}{2}[A_{\alpha\lambda}(0)A_{\beta\lambda'}(\tau)-A_{\alpha\lambda}(\tau)A_{\beta\lambda'}(0)] \right. \nonumber\\
\left.
+\frac{1}{2}[A_{\beta\lambda'}(0)+A_{\beta\lambda'}(\tau)]\bar\sigma_{\alpha\lambda}(0)
+ [A_{\alpha\lambda}(0)-A_{\alpha\lambda}(\tau)]\left.\sum_{n=0}^\infty \right.'
\bar\sigma_{\beta\lambda'}(n\tau)\right\rangle 
\end{eqnarray}
and 
\begin{equation} 
\label{SGT2}
\frac{\tau}{N \,k_B T}\left.\sum_{n=0}^\infty \right.'
\langle\hat k_\lambda \bar\sigma_{\alpha\lambda}(0)\vert 
\hat k_{\lambda'}\bar\sigma_{\beta\lambda'}(n \tau)\rangle\ .  
\end{equation} 
Stationarity implies that the first term in  
Eq. (\ref{SGT1}) equals zero, and that the last term reduces to 
\begin{equation} 
[A_{\alpha\lambda}(0)-A_{\alpha\lambda}(\tau)]\left.\sum_{n=0}^\infty \right.'
\bar\sigma_{\beta\lambda'}(n \tau) = 
-\frac{1}{2}[A_{\alpha\lambda}(\tau) + A_{\alpha\lambda}(0)]
\bar\sigma_{\beta\lambda'}(0).  
\end{equation}  
Stationarity and time reversal invariance imply that the remaining term also 
vanishes. Alternatively, the explicit form of $A_{\alpha\lambda}$, 
Eq (\ref{A_def}), can be used to show that this term vanishes. 
The expression in Eq. (\ref{SGT1}) is therefore zero, so that the 
Green-Kubo relation for the viscous transport coefficients  
is still given by (\ref{VTCGK}), 
but with the stress tensor $\bar\sigma_{\alpha\lambda}$.  

A similar calculation shows that the thermal diffusivity is given by 
(\ref{DT}), with $\chi_\lambda(n\tau)$ replaced by  
$\bar\chi_\lambda(n\tau)\equiv\bar\chi^{kin}_\lambda(n\tau)+
\bar\chi^{rot}_\lambda(n\tau)$,  
with 
\begin{equation} 
\label{VIS3N}
\bar\chi^{kin}_\lambda(n\tau) =-\sum_j[-\frac{1}{2}v_j^2(n\tau) 
v_{j\lambda}(n\tau) + 
                                 k_BTv_{j\lambda}(n\tau)], \ \ \ \ \
\bar\chi^{rot}_\lambda(n\tau) =\frac{1}{2\tau}\sum_j v_j^2(n\tau)
B_{j\lambda}(n\tau).  
\end{equation} 
Note that the new stress tensors do not depend on $\xi$, the space-fixed 
cell coordinates of the particles.   

An alternative way to derive these results is to note that time reversal invariance 
can be used to rewrite (\ref{VTCGK}) and (\ref{DT}) as sums from $-\infty$ 
to $+\infty$. In this way, the discussion in the preceeding paragraph 
of the $n=0$ term can be avoided. 

\section{Correlations involving {\bf B}'s } 

$B_{i\alpha} (n\tau)$  
is the $\alpha$-component of the difference 
between the change in the shifted cell coordinates during one streaming  
step and the actual distance traveled, $\tau v_{i\alpha}$. It has 
a number of important properties which simplify the calculation of the 
transport coefficients. In particular, it will be shown that all 
stress-stress correlation functions involving one {\bf B} in the 
GK relations for the transport coefficients are zero, so that, for 
example, $\Lambda_{\alpha\beta}(\nu_1,\nu_2,\gamma,\epsilon;\hat{\bf k}) =  
\Lambda^{kin}_{\alpha\beta}(\nu_1,\nu_2,\gamma,\epsilon;\hat{\bf k}) + 
\Lambda^{rot}_{\alpha\beta}(\nu_1,\nu_2,\gamma,\epsilon;\hat{\bf k})$, with   
\begin{equation} 
\label{dec1}
\Lambda^{kin}_{\alpha\beta}(\nu_1,\nu_2,\gamma,\epsilon;\hat{\bf k}) \equiv 
\frac{\tau}{Nk_BT}\left.\sum_{n=0}^\infty\right.' 
\langle\hat k_\lambda \bar\sigma^{kin}_{\alpha\lambda}(0)\vert 
\hat k_{\lambda'} \bar\sigma^{kin}_{\beta\lambda'}(n\tau)\rangle  
\end{equation} 
and 
\begin{equation} 
\label{dec2}
\Lambda^{rot}_{\alpha\beta}(\nu_1,\nu_2,\gamma,\epsilon;\hat{\bf k}) \equiv 
\frac{\tau}{Nk_BT}\left.\sum_{n=0}^\infty\right.' 
\langle\hat k_\lambda \bar\sigma^{rot}_{\alpha\lambda}(0)\vert 
\hat k_{\lambda'} \bar\sigma^{rot}_{\beta\lambda'}(n\tau)\rangle .   
\end{equation} 

The kinetic contributions to the viscosity were 
calculated previously in both 2d~\cite{ihle_03b} and 3d~\cite{tuze_03}, and 
will not be discussed here.  
Properties of the {\bf B}-correlations which enable an explicit 
evaluation of expression (\ref{dec2}) are derived in the following 
subsections. 

\subsection{Factorization of $\bf{B}\!\!-\!\!\bf{v}$ correlations} 
\label{subsec_ac}

The first one, $\langle B_{i\alpha}(n\tau)\rangle=0$ for arbitrary $n$, 
implies that on average, the distance traveled by a particle during one 
time step is the average of difference of the shifted cell coordinates before 
and after the streaming step.  This can be shown as follows. Consider 
\begin{equation} 
\label{B_av}
\langle B_{ix}(0)\rangle = \langle \xi_{ix}^s(\tau)-\xi_{ix}^s(0)
                         - \tau v_{ix}(0)\rangle\;\;.
\end{equation}  
The ensemble average includes averaging over the initial coordinates and 
velocities of all particles, as well as averages over the shift and collision 
matrix at each time step. Without loss of generality, assume that at 
$t=0$, the $x$-coordinate of particle $i$ is in the interval $[0,a)$. For 
$na\le X\equiv x_i(0)+\tau v_{ix}(0)<(n+1)a$, the average of  
$\xi_{ix}^s(\tau)\equiv \xi_{ix}(\tau)-\Delta \xi_{ix}^s(0)$ over the 
random shift $\delta$ at time $\tau$, denoted by $\langle ~\rangle_{\delta_\tau}$,
at fixed particle coordinate and velocity, is (see Eq. (31) of \cite{ihle_03b})    
\begin{equation} 
\langle \xi_{ix}^s(\tau)\vert_X\rangle_{\delta_\tau} = na -  
\langle \Delta\xi_{ix}^s\vert_X\rangle_{\delta_\tau} = X-a/2 , 
\end{equation} 
so that 
\begin{equation} 
\langle B_{ix}(0)\rangle = \langle -a/2+x_i(0)- 
                           \xi_{ix}^s(0)\rangle\;\;.
\end{equation} 
Finally, averaging over the shift at $t=0$ gives 
$\langle \xi_{ix}^s(0)\vert_x\rangle_{\delta_0} = -a/2+x_i(0)$, so that 
$\langle B_{ix}(0)\rangle = 0 $. 

Similar arguments can be used to show that the cross terms 
in the GK expressions for the transport coefficients involving one $B$ 
are zero. For the shear viscosity, these terms involve 
correlations of the form $\langle v_{jx}(n\tau)v_{jy}(n\tau)  
v_{ix}(m\tau)B_{iy}(m\tau)\rangle$. Consider first the case $n=m=0$. 
Performing the average over the shift $\delta$ at $t=\tau$, the average 
reduces to 
\begin{equation} 
\label{AV1}
\langle v_{jx}(0)v_{jy}(0)v_{ix}(0) [y_i(0)-a/2-\xi_{iy}^s(0)]\rangle. 
\end{equation}  
The average over the shift at $t=0$ gives zero because it doesn't affect 
the particle's initial velocities or positions. 
Consider now $m=0$ and $n=1$. In this case, perform 
first the average over the random shift at $t=0$. The result is 
\begin{equation} 
\label{ex1} 
\langle v_{jx}(\tau)v_{jy}(\tau)v_{ix}(0)[\xi_{iy}^s(\tau)+a/2-y_i(\tau)]
\rangle  \;\;.
\end{equation} 
If the probability of any given configuration at $t=\tau$ in a shifted cell 
containing 
particle $i$ is independent of $\delta$, the average over the shift at 
$t=\tau$ factorizes. This is, in fact, the case since the average in 
(\ref{ex1}) entails an integration over the initial particle coordinates 
and velocities at $t=0$. In this case, the average over $\delta$ can be 
performed; since $\langle \xi_{iy}^s(\tau)+a/2-y_i(\tau)\rangle\vert_\delta$ 
vanishes, the result of this averaging is zero. An alternative,  
more detailed discussion of this proof is given in the Appendix. 

The argument for general $m$ and $n$ is similar. Analogous reasoning can 
be used to show that correlations such as 
\begin{equation} 
\label{factorize}
\langle v_{ix}(0)v_{jx}(n\tau)B_{iy}(0)B_{jy}(n\tau)\rangle =   
\langle v_{ix}(0)v_{jx}(n\tau)\rangle\langle B_{iy}(0)B_{jy}(n\tau)\rangle   
\end{equation} 
factorize for arbitrary $n$. 

\subsection{Autocorrelation of $\bf{B}$'s}  

It is straightforward to evaluate equal-time correlation functions of the 
$B$ variables. Using the results derived in the Appendix and Eq. (36) of 
\cite{ihle_03b}, one has 
\begin{equation} 
\langle B_{ix}^2(0)\rangle = a^2/3 \ \ \ {\rm and} \ \ \ \ 
\langle B_{ix}(0)B_{jx}(0)\rangle = a^2/6, \ \ \ {\rm for} \ i\neq j.  
\end{equation}  

Correlation functions such as $\langle B_{ix}(0)B_{ix}(\tau)\rangle$ can 
be evaluated as follows. Take $0\le x_i(0)<a$, 
$\Delta\xi_{ix}\equiv\Delta\xi_{ix}(0)=ma$, and $\Delta\xi_{ix}(\tau)=na$. 
Here and in the following expressions the argument $(0)$ will be omitted for clarity.
Averaging over 
the random shift $\delta$ at $t=2\tau$,   
\begin{eqnarray} 
\langle B_{ix}(0)B_{ix}(\tau)\rangle &=& 
-\langle [(m+1/2)a-x_i(\tau)] 
[\Delta\xi_{ix}+\Delta\xi^s_{ix}(-\tau)-\Delta\xi^s_{ix}-\tau v_{ix}]\rangle 
\nonumber \\ 
&+& \langle\Delta\xi^s_{ix}
[\Delta\xi_{ix}+\Delta\xi^s_{ix}(-\tau)-\Delta\xi^s_{ix}-\tau v_{ix}]\rangle. 
\label{BBt1} 
\end{eqnarray} 
Since $(m+1/2)a-x_i(\tau)$ does not depend on random shifts at time $0$ and 
$\tau$, the average over $\delta_0$ and $\delta_\tau$ in the first term on the 
right hand side of Eq. (\ref{BBt1}) at fixed $x_i(0)$ and $v_{ix}(0)$ vanishes, 
so that  
\begin{equation} 
\langle B_{ix}(0)B_{ix}(\tau)\rangle = 
-\langle(\Delta\xi^s_{ix})^2\rangle + \langle\Delta\xi^s_{ix}
[\Delta\xi_{ix}+\Delta\xi^s_{ix}(-\tau)-\tau v_{ix}]\rangle.   
\end{equation} 
Averaging the second term on the right hand side of this equation over 
$\delta_0$ and $\delta_\tau$, one finds 
\begin{equation} 
\label{BBt2} 
\langle\Delta\xi^s_{ix}[\Delta\xi_{ix}+
          \Delta\xi^s_{ix}(-\tau)-\tau v_{ix}]\rangle = 
\frac{1}{a}\int_0^a dx_i \sum_{m=-\infty}^\infty \int_{(ma-x_i)/\tau}^  
{[(m+1)a-x_i]/\tau}[(m+1/2)a-x_i-\tau v_{ix}]^2 w(v_{ix})dv_{ix}. 
\end{equation} 
Comparing with Eqs. (18) and (32) of \cite{ihle_03b}, 
it can be shown that (\ref{BBt2}) 
is equal to $-\langle(\Delta\xi_{ix})^2\rangle+2
\langle\Delta\xi_{ix}\Delta\xi^s_{ix}\rangle + \lambda^2 + a^2/12=a^2/12$, 
where the last equality follows from
\begin{equation}
\langle\Delta\xi_{ix}\Delta\xi^s_{ix}\rangle
=\frac{1}{2} \left[ \langle \Delta\xi_{ix}^2 \rangle -\lambda^2\right]
\end{equation}
given as Eq. (36) in \cite{ihle_03b}. 

Finally, using 
(\ref{DXI2}), 
\begin{equation}
\langle B_{ix}(0)B_{ix}(\tau)\rangle = -a^2/6.  
\end{equation}   
The average $\langle B_{ix}(0)B_{jx}(\tau)\rangle$ can be evaluated in 
a similar fashion. Take $m_0a\le x_i(0)<(m_0+1)a$, $n_0a\le x_j(0)< (n_0+1)a$, 
$\Delta\xi_{ix}(0)=m_1a$, $\Delta\xi_{jx}(0)=n_1a$, 
$\Delta\xi_{ix}(\tau)=m_2a$ and $\Delta\xi_{jx}(\tau)=n_2a$. Averaging over 
$\delta_{2\tau}$, one has 
\begin{eqnarray} 
\langle B_{ix}(0)B_{jx}(\tau)\rangle &=& 
-\langle [(n_0+n_1+1/2)a-x_j(\tau)] 
[\Delta\xi_{ix}+\Delta\xi^s_{ix}(-\tau)-\Delta\xi^s_{ix}-\tau v_{ix}]\rangle 
\nonumber \\ 
&+& \langle\Delta\xi^s_{jx}
[\Delta\xi_{ix}+\Delta\xi^s_{ix}(-\tau)-\Delta\xi^s_{ix}-\tau v_{ix}]\rangle. 
\label{BBt3} 
\end{eqnarray} 
Again, since $(n_0+n_1+1/2)a-x_i(\tau)$ does not depend on random shifts at 
time $0$ and 
$\tau$, the average over $\delta_0$ and $\delta_\tau$ in the first term on the 
right hand side of Eq. (\ref{BBt3}) vanishes, 
so that  
\begin{equation} 
\langle B_{ix}(0)B_{jx}(\tau)\rangle = 
-\langle\Delta\xi^s_{ix}\Delta\xi^s_{jx}\rangle + \langle\Delta\xi^s_{jx}
[\Delta\xi_{ix}+\Delta\xi^s_{ix}(-\tau)-\tau v_{ix}]\rangle.   
\end{equation} 
Using Eqs. (\ref{DXIJ}), (\ref{a1}), (\ref{a2}), and (\ref{a4}) from Appendix, we have for $i\neq j$   
\begin{equation} 
\langle B_{ix}(0)B_{jx}(\tau)\rangle = -a^2/12. 
\end{equation}
 
All $\bf{B}$-correlation functions for time lags greater than $\tau$ are zero. 
To understand this, consider $\langle B_{ix}(0)B_{jx}(2\tau)\rangle$. 
Averaging over $\delta_{3\tau}$, 
\begin{equation} 
\label{BB2t}
\langle B_{ix}(0)B_{jx}(2\tau)\rangle = 
\langle[\Delta_{jx}(\tau)-(m+1/2)a+x_j(2\tau)]
[\Delta\xi_{ix}+\Delta\xi^s_{ix}(-\tau)-\tau v_{ix}]\rangle, 
\end{equation} 
where $ma$ is the cell coordinate of particle $j$ at $t=2\tau$. The second 
term in (\ref{BB2t}) has no dependence on $\delta_{2\tau}$, while average of 
the first term gives zero. Again, this requires that the probability 
of any given configuration in a shifted cell is independent of 
$\delta_{2\tau}$.  

\begin{figure}
\begin{center}
\vspace{2cm}
\includegraphics[width=5in,angle=0]{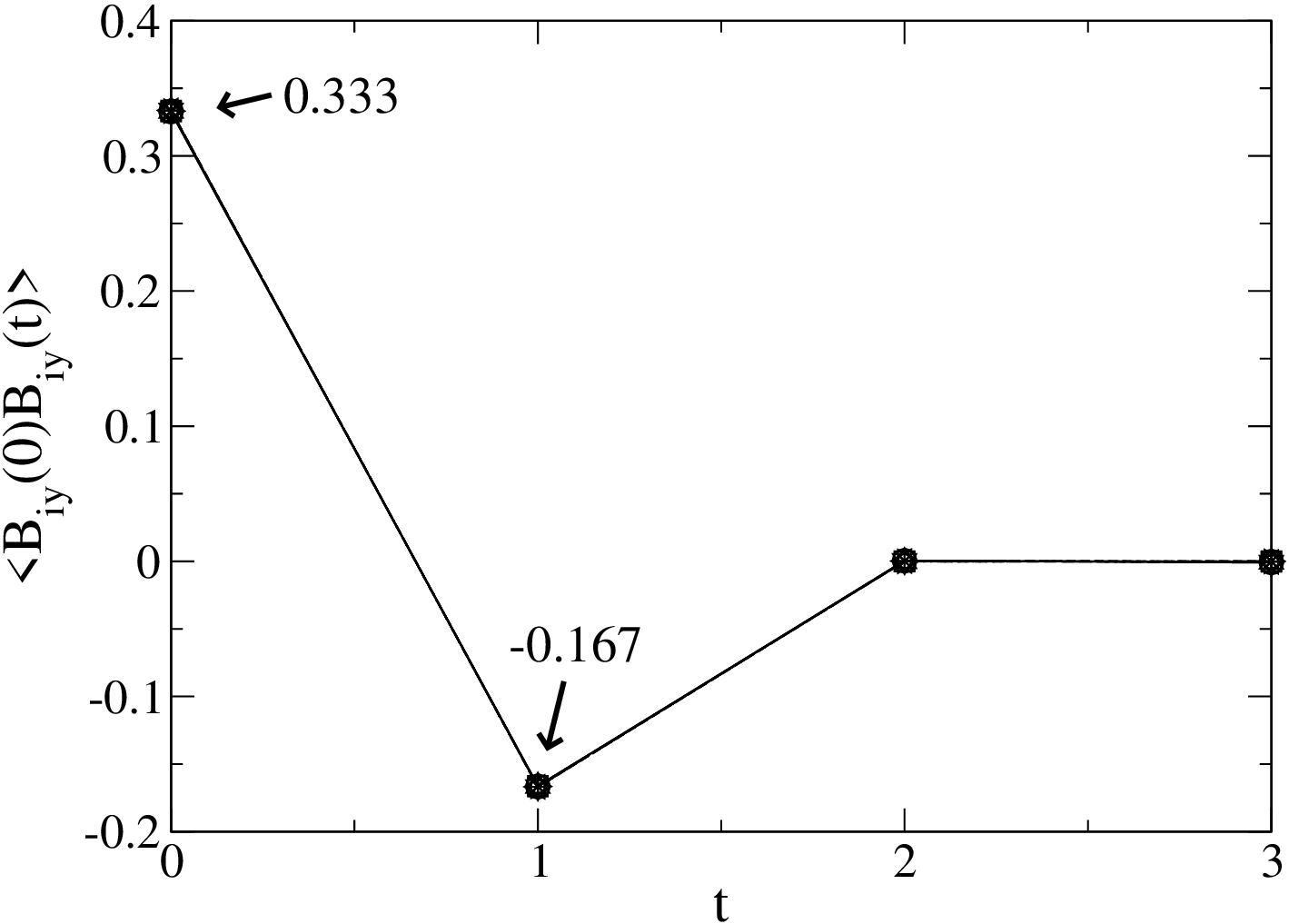}
\caption{$\langle B_{iy}(0)B_{iy}(t)\rangle$ as a function of time. 
Results for $\lambda/a=0.05$, 0.10, 0.50,1.00, and for collision
angles $\alpha=60^\circ$, $90^\circ$, and $120^\circ$ are plotted, indicating 
that there is no dependence on the value of the mean free path.  
Time averages over $10^6$ iterations were used to obtain the data. 
Parameters: $L/a=32$ and $M=5$.}
\label{fig_Byii}
\end{center}
\end{figure}

\begin{figure}
\begin{center}
\includegraphics[width=5in,angle=0]{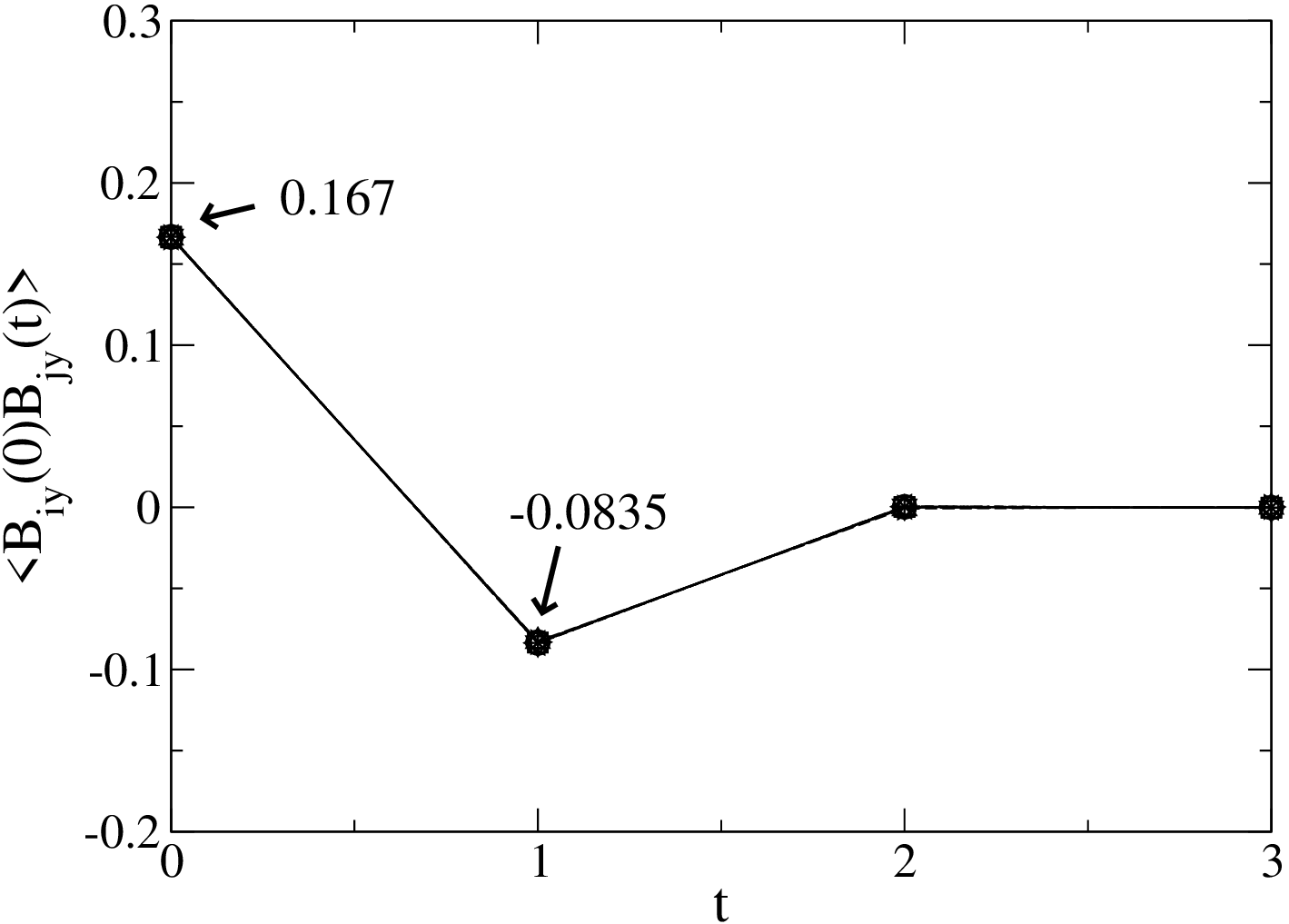}
\caption{$\langle B_{iy}(0)B_{jy}(t)\rangle$ as a function of time. 
Identical results are obtained for $\lambda=0.05, 0.10, 0.50,1.00$ and for 
collision angles $\alpha=60^\circ$, $90^\circ$, and $120^\circ$. 
Time averages over $10^6$ iterations 
were used to obtain the data. Parameters: $L/a=32$ and $M=5$.}
\label{fig_Byij}
\end{center}
\end{figure}

\begin{figure}
\begin{center}
\vspace{2cm}
\includegraphics[width=5in,angle=0]{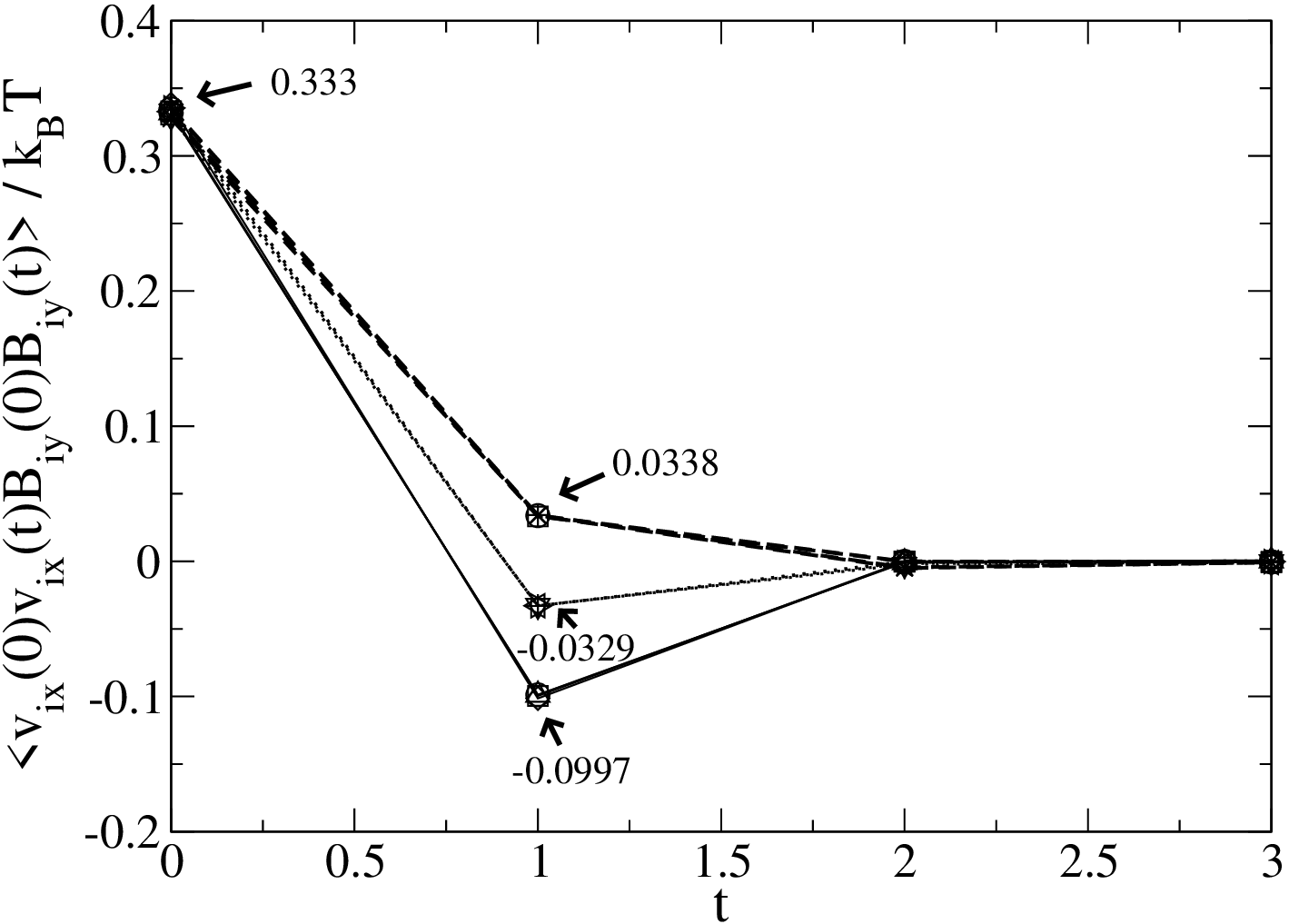}
\caption{$\langle v_{ix}(0)v_{ix}(t)B_{iy}(0)B_{iy}(t)\rangle$ as a 
function of time.
Identical results are obtained for $\lambda=0.05, 0.10, 0.50$, and 1.00.
Solid, dotted and dashed lines correspond to collision angles 
$\alpha=60^\circ$, $90^\circ$, and $120^\circ$, respectively. 
Time averages over $10^6$ iterations were used to obtain the data. 
Parameters: $L/a=32$ and $M=5$.}
\label{fig_vxByii}
\end{center}
\end{figure}

These results can be summarized by the relation   
\begin{equation}
\label{VIS8}
\langle B_{i \alpha}(n\tau) B_{j \beta}(m\tau) \rangle=\frac{a^2}{12}
\, \delta_{\alpha\beta} (1+\delta_{ij})
\left[2 \delta_{n,m} - \delta_{n,m+1}-\delta_{n,m-1}  \right] . 
\end{equation}
Fig. \ref{fig_Byii} presents simulation results for $\langle B_{iy}(0)
B_{iy}(t)\rangle$ in $d=2$ for various collision angles $\alpha$ and 
a range of mean-free paths. Fig. \ref{fig_Byij} contains corresponding 
results for $\langle B_{iy}(0)B_{jy}(t)\rangle$. In both cases, the agreement 
with result (\ref{VIS8}) is excellent. Simulation results for 
$\langle v_{ix}(0)v_{ix}(t)B_{iy}(0)B_{iy}(t)\rangle$ as a function 
of time are presented in Fig. \ref{fig_vxByii} for a similar range of 
parameters; the results are in agreement with the prediction of 
subsection \ref{subsec_ac} that this autocorrelation function factorizes, 
and that the resulting $B$-correlations are given by Eq. (\ref{VIS8}).  
 
It follows that there are only two---a pure kinetic and a 
pure rotational (or collision)---contributions to the transport coefficients. 
Relation (\ref{VIS8}) is of central importance, because it contains all the 
geometrical features of the grid that contribute to the transport 
coefficients, and is independant of specific collision rules and particle
properties. Since the kinetic contribution to the stress tensor is symmetric 
and has been calculated elsewhere, we concentrate here of the (collisional) 
contributions arising from $\bf{B}$-correlations. 

\subsection{Viscosities} 

Explicit expressions for the collisional contributions to the 
viscous transport coefficients can be obtained by considering various 
choices for ${\bf \hat k}$ and $\alpha$ and $\beta$ in Eq. (\ref{dec2}) 
and using (\ref{ST}). Taking 
${\bf \hat k}$ in the $y$-direction and $\alpha=\beta=1$ yields
\begin{equation} 
\label{r1}
\nu^{rot}\equiv\nu_1^{rot}+\nu_2^{rot} =     
\frac{1}{\tau Nk_BT}\left.\sum_{n=0}^\infty\right.'   
\sum_{i,j} \langle v_{ix}(0)B_{iy}(0)v_{ix}(t)B_{iy}(n\tau)\rangle. 
\end{equation} 
Eq. (\ref{r1}) is the expression used in Ref. \cite{tuze_04} to determine the 
collisional contribution to the shear viscosity. 
$\nu_1^{rot}$ and $\nu_2^{rot}$ are the viscosities associated with the 
symmetric and the anti-symmetric contributions to the matrix of viscous 
transport coefficients. 

Other choices for ${\bf \hat k}$ and $\alpha$ and $\beta$ 
yield expressions for other linear combinations of the transport coefficients. 
In particular, the choice ${\bf \hat k}=(1,0,0)$ and $\alpha=\beta=1$ 
yields 
\begin{equation} 
\label{r3} 
(1+(d-2)/d)\nu_1^{rot}+\gamma^{rot}+\epsilon^{rot} = 
\frac{1}{\tau Nk_BT}\left.\sum_{n=0}^\infty\right.'
\sum_{i,j} \langle v_{ix}(0)B_{ix}(0)v_{ix}(t)B_{ix}(n\tau)\rangle.
\end{equation}
However, because of (\ref{VIS8}), the right hand side of (\ref{r3}) is 
equal to (\ref{r1}), so that  
\begin{equation} 
(1+(d-2)/d)\nu_1^{rot}+\gamma^{rot}+\epsilon^{rot} = \nu^{rot} .  
\end{equation}
Finally, for $\hat k=(1,1,0)/\sqrt{2}$ and $\alpha=1$, $\beta=2$, one has 
\begin{equation} 
\label{r2} 
((d-2)/d)\nu_1^{rot}-\nu_2^{rot}+\gamma^{rot}=0, 
\end{equation}
since the resulting stress-stress correlation functions are zero. 
These results imply that $\epsilon^{rot}=0$, and that the longitudinal 
component of (\ref{ST}), which is the viscous contribution to the sound 
attenuation, is $\nu^{rot}$.  
Finally, using these results in (\ref{ST}), it follows that the collision 
contribution to the macroscopic stress tensor is 
\begin{equation} 
\label{ST1} 
\hat\sigma^{rot}_{\alpha\beta} = (\nu_1^{rot}+\nu_2^{rot})
 \partial_\beta v_\alpha = \nu^{rot} \partial_\beta v_\alpha 
\end{equation} 
up to a tensor $\tensor{G}$ with vanishing divergence,  
$\partial_\beta G_{\alpha\beta}=0$, which therefore will not appear in 
the linearized hydrodynamic equations. The collisional contribution to the 
effective shear viscosity is therefore $\nu^{rot}$, and the   
viscous contribution to the sound attenuation 
is also $\nu^{rot}$, instead of the standard result, 
$2(d-1)\nu/d+\gamma$, for simple isotropic fluids. 
The corresponding hydrodynamic equation for the momentum density is 
therefore  
\begin{equation}\label{mom_eq}  
\rho\left[\frac{\partial{\bf v}}{\partial t}+({\bf v}\cdot\nabla){\bf v}
\right] = -\nabla p + \rho(\nu^{kin}+\nu^{rot})\Delta{\bf v}
+ \frac{2-d}{d}\nu^{kin}\nabla (\nabla\cdot{\bf v}), 
\end{equation} 
where we have used the fact that the kinetic contribution to the microscopic 
stress tensor, $\bar\sigma^{kin}$ in (\ref{VIS2N}), is symmetric, and 
$\gamma^{kin}=0$ \cite{ihle_03b}. Note that there is no collisional 
contribution to the last term in Eq. (\ref{mom_eq}). For $d=2$, the viscous 
contribution to the sound attenuation coefficient in a simple liquid is 
$\nu+\gamma$. The results of the current calculation are consistent with 
this result, with $\nu=\nu^{kin}+\nu^{rot}$ and $\gamma=0$ (as expected 
for a fluid with an ideal gas equation of state). In $d=3$, while the shear 
viscosity is still given by $\nu=\nu^{kin}+\nu^{rot}$, the viscous 
contribution to the sound attenuation coefficient is $4\nu^{kin}/3+\nu^{rot}$, 
instead of $4(\nu^{kin}+\nu^{rot})/3 + \gamma$. 
The sound attenuation coefficient of the SRD model in three dimensions is 
therefore slightly smaller than in a simple liquid.
Note, however, that the hydrodynamic equations are not affected.  
Although only certain linear combinations of the collisional contributions 
to the viscous transport coefficients are determined by Eqs. 
(\ref{r1})--(\ref{r2}), all coefficients in
the linearized hydrodynamic equations are uniquely determined.

Equilibrium measurements of the time-dependent density correlations 
\cite{tuze_sound} in two dimensions yield results for the sound attenuation 
coefficient which are in good agreement with the theoretical predictions. A 
detailed comparison of these simulation results with theory will be presented 
elsewhere~\cite{tuze_sound}. Currently, there are no similar measurements in three dimensions. 

The fact that the entropy of a fluid increases as a result of irreversible 
processes leads to certain positivity conditions on the transport coefficients
\cite{land_59}. SRD obeys an H-theorem \cite{male_99, ihle_03a}, which 
implies that the entropy production is always non-negative. In Appendix E, 
it is shown, using a generalization of an argument from \cite{land_59}, that  
the requirement of a positive entropy production leads to the conditions 
$\nu^{kin}+\nu_1^{rot} \geq 0$, $\nu_2^{rot} \geq 0$, and
$\gamma \geq 0$. Note that the result of Pooley and Yeomans \cite{pool_05} 
for the collisional stress tensor 
$\hat{\sigma}_{\alpha \beta}^{rot}=\nu^{rot}\,\partial_\beta v_\alpha$
amounts to assuming $\nu_1^{rot}=\nu_2^{rot}=\nu^{rot}/2$ and 
$\gamma=\nu^{rot}/d$.

\section{Explicit expressions for the collisional contributions to the 
transport coefficients}

\subsection{Viscosities} 

Using the results of the previous sections, the collisional contribution 
to the viscosity can be written as  
\begin{equation} 
\label{VIS10}
\nu^{rot}=\nu_1^{rot}+\nu_2^{rot}  = \frac{1}{2\tau N \,k_B T}  \sum_{i,j=1}^N 
\left\{
\langle v_{ix}(0) v_{jx}(0) \rangle \langle B_{iy}(0) B_{jy}(0) \rangle +
2 \langle v_{ix}(0) v_{jx}(\tau) \rangle \langle B_{iy}(0) B_{jy}(\tau) 
\rangle \right\} \;.
\end{equation}

\begin{figure}
\begin{center}
\vspace{2cm}
\includegraphics[width=5in,angle=0]{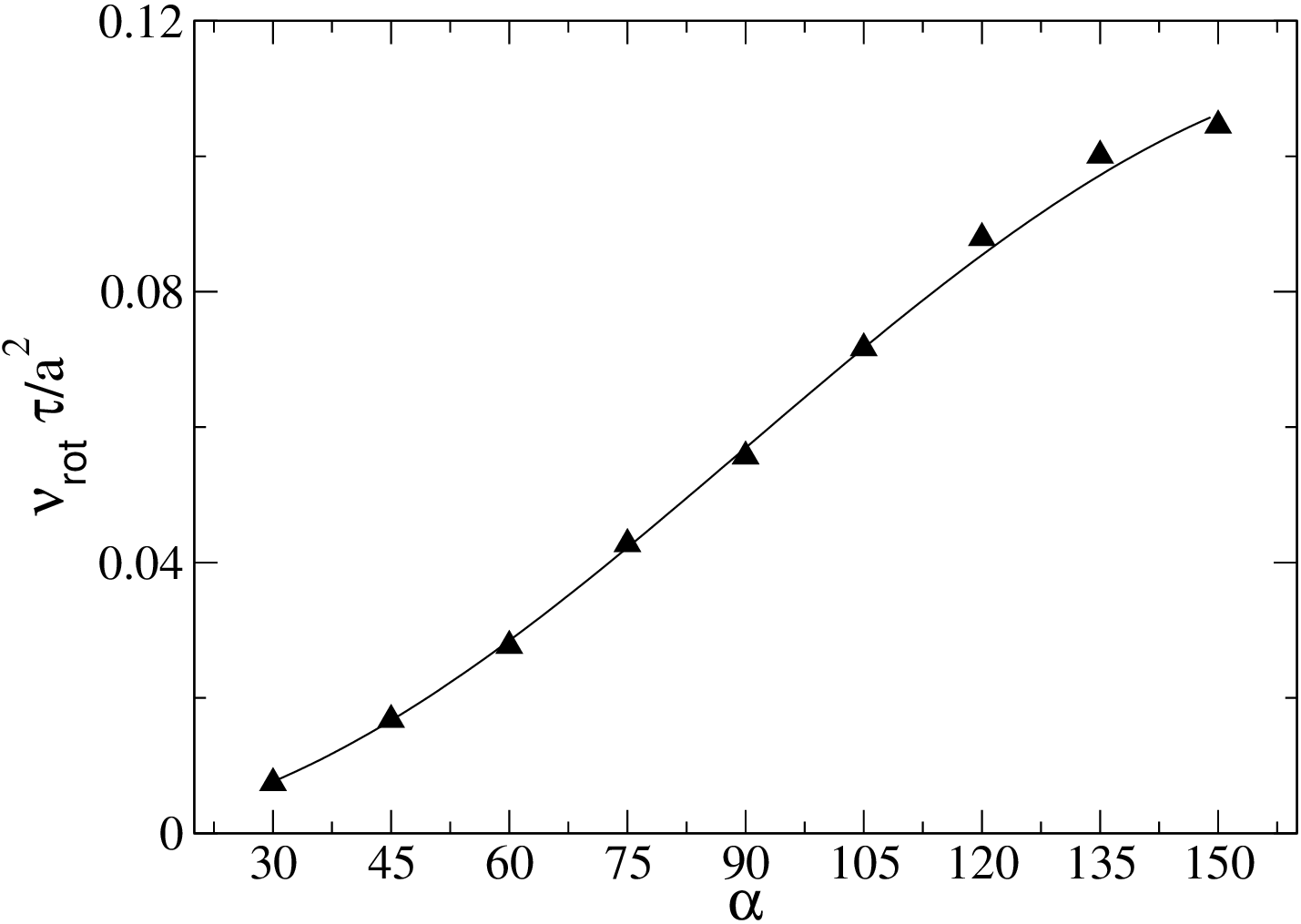}
\caption{The normalized collisional contribution to the kinematic viscosity, 
$\nu^{rot}\tau/a^2$, as a function of collision angle $\alpha$. 
The solid line is the theoretical prediction (\ref{VIS11}). 
The data were obtained by time averaging over 360,000 iterations.
Parameters: $L/a=16$, $\lambda/a=0.1$, $M=3$ and $\tau=1$.}
\label{fig_nurot}
\end{center}
\end{figure}

It is straightforward to evaluate the various contributions 
to the right hand side of (\ref{VIS10}). In particular, note that since 
velocity correlation functions  
only at equal time and for a time lag of one time step are required, 
molecular chaos can be assumed when evaluating these contributions, since 
it was shown in \cite{tuze_04} that additional correlation effects only 
occur for larger time lags. Using (\ref{VIS8}), the first term on the 
right hand side of Eq. (\ref{VIS10}) reduces to 
\begin{equation}\label{exp1}  
\sum_{i,j=1}^N \langle B_{iy}(0) B_{jy}(0) \rangle
\langle v_{ix}(0) v_{jx}(0) \rangle  = \frac{a^2}{3}
\sum_{i} \langle v_{ix}^2(0)\rangle + \frac{a^2}{6}\sum_i\sum_{j\ne i} 
\langle v_{ix}(0)v_{jx}(0)\rangle\,. 
\end{equation} 
Momentum conservation, namely $\sum_k v_{kx}(0)=0$, can be used to write 
\begin{equation} 
\sum_i\sum_{j\ne i} \langle v_{ix}(0)v_{jx}(0)\rangle = -
\sum_i \langle v_{ix}^2(0)\rangle\,, 
\end{equation} 
so that the right hand side of (\ref{exp1}) reduces to 
\begin{equation} 
\frac{a^2}{6}\sum_i\langle v_{ix}^2(0)\rangle = \frac{a^2}{6}N_f k_BT\,, 
\end{equation} 
where $N_f=N-1$, because of momentum conservation. Similarly, using 
momentum conservation at $t=\tau$, the second term in on the right hand 
side of (\ref{VIS10}) reduces to 
\begin{equation}  
\sum_{i,j=1}^N \langle B_{iy}(0) B_{jy}(\tau) \rangle
\langle v_{ix}(0) v_{jx}(\tau) \rangle  = -\frac{a^2}{12}\sum_i 
\langle v_{ix}(0)v_{ix}(\tau)\rangle\,.
\end{equation} 
It follows that 
\begin{equation}\label{res1} 
\nu^{rot} = \frac{a^2}{12\tau}\left\{1-\frac{\langle v_{ix}(0)v_{ix}(\tau)
\rangle}{k_BT}\right\} + O(1/N). 
\end{equation} 
For $d=2$, if there are $m_i$ particles in the collision cell 
$\xi_i^s(\tau)$, the ensemble average of the term in brackets in 
(\ref{res1}) is 
\begin{equation} 
1-\frac{1}{k_BT}\langle v_{ix}(0)v_{ix}(\tau)\rangle\vert_{m_i} = 
(1-1/m_i)[1-\cos(\alpha)]\,, 
\end{equation} 
where $\alpha$ is the collision angle. In three dimensions,
the corresponding expression is  
\begin{equation} 
1-\frac{1}{k_BT}\langle v_{ix}(0)v_{ix}(\tau)\rangle\vert_{m_i} = 
\frac{2}{3}(1-1/m_i)[1-\cos(\alpha)]\,,
\end{equation} 
for both models A and B.   
To obtain the final result, we now need to average over the number 
of particles, $m_i$, in the collision cell. If the average number of 
particles per cell is $M$, the probability that there are $m_i$ particles 
in cell $\xi^s_i$ is given by the Poisson distribution $P_p(m_i,M)={\rm e}^{-M} 
M^{m_i}/m_i!$. The corresponding (normalized) probability that a given 
particle, $i$, is in a cell containing a total number of particles 
$m_i$ is $m_iP_p(m_i,M)/M$. Averaging now over the  
number of particles in a cell, we have, finally,  
\begin{equation}
\label{VIS11} 
\nu^{rot}=\frac{a^2}{6 d \tau} \left(\frac{M-1+e^{-M}}{M}\right)
[1-\cos(\alpha)] \;,
\end{equation}
for all the collision models we considered (the standard model in $d=2$ 
and both models A and B in $3d$).
Eq. (\ref{VIS11}) agrees with the result of Refs. \cite{kiku_03} and 
\cite{pool_05} obtained using a completely different non-equilibrium 
approach in shear flow. 
Simulation results for the rotational contribution to the viscosity,
Eq. (\ref{VIS11}), are compared with the theoretical prediction 
in Fig. \ref{fig_nurot} for small $M=3$, where our earlier approximation 
\cite{ihle_01} for $M\gg 1$ would not be accurate.  
The new expression correctly describes the limit $M\rightarrow 0$, 
where the collisional viscosity should vanish. 

\subsection{Thermal diffusivity}

The collisional contribution to the thermal diffusivity can be calculated 
in a similar fashion. In particular, taking $\hat{\bf k}=\hat x$ 
in (\ref{DT}) and (\ref{DT1}), and using (\ref{VIS3N}), we have 
$D_T=D_T^{kin}+D_T^{rot}$ with
\begin{eqnarray}
\label{DTkin}
D_T^{kin} & = & \frac{\tau}{c_p N \,k_B T^2}
\left.\sum_{n=0}^\infty \right.'
\sum_{i,j=1}^N \langle 
\left[{v_i^2(0) / 2} - k_B T\right]\left[{v_j^2(n \tau) / 2} -k_B T\right]
v_{ix}(0) v_{jx}(n \tau) \rangle \\
\label{DTrot}
D_T^{rot} & = & \frac{1}{8 c_p \tau N \,k_B T^2}  \sum_{i,j=1}^N
\left\{
\langle v_i^2(0) v_j^2(0) \rangle \langle B_{ix}(0) B_{jx}(0) \rangle +
2 \langle v_i^2(0) v_j^2(\tau) \rangle \langle B_{ix}(0) B_{jx}(\tau)
\rangle \right\} \;.
\end{eqnarray}
The kinetic contributions to the thermal diffusivity were 
calculated previously in both 2d~\cite{ihle_03b} and 3d~\cite{tuze_03}. 
Using the results presented earlier in this paper, it is straightforward 
to evaluate the collisional contribution to the thermal diffusivity. 
Just as momentum conservation was used to simplify the 
calculation of the collisional contribution to the viscosity, energy 
conservation, $\sum_{k=1}^N v_k^2={\rm const.}$, and (\ref{VIS8}) can be 
used to show that (\ref{DTrot}) reduces to    
\begin{equation}\label{res2}   
D_T^{rot}  =  \frac{a^2d}{24 \tau } 
\left\{1 - \frac{\langle v_i^2(0) v_i^2(\tau) \rangle}{d(d+2)(k_BT)^2} 
\right\} \;.
\end{equation} 

For $d=2$, if there are $m_i$ particles in the collision cell 
$\xi_i^s(\tau)$, the ensemble average of the term in brackets in 
(\ref{res2}) is 
\begin{equation} 
\left.1-\frac{\langle v^2_i(0)v^2_i(\tau)\rangle}{8(k_BT)^2}
\right\vert_{m_i} = \frac{1}{m_i}\left(1-\frac{1}{m_i}\right)
[1-\cos(\alpha)]\,. 
\end{equation} 
In three dimensions, the corresponding expression is  
\begin{equation} 
\left.1-\frac{\langle v^2_i(0)v^2_i(\tau)\rangle}{15(k_BT)^2}
\right\vert_{m_i} = \frac{8}{15 m_i}\left(1-\frac{1}{m_i}\right)
[1-\cos(\alpha)]\,,
\end{equation} 
for both models A and B.   

\begin{figure}
\begin{center}
\vspace{2cm}
\includegraphics[width=5in,angle=0]{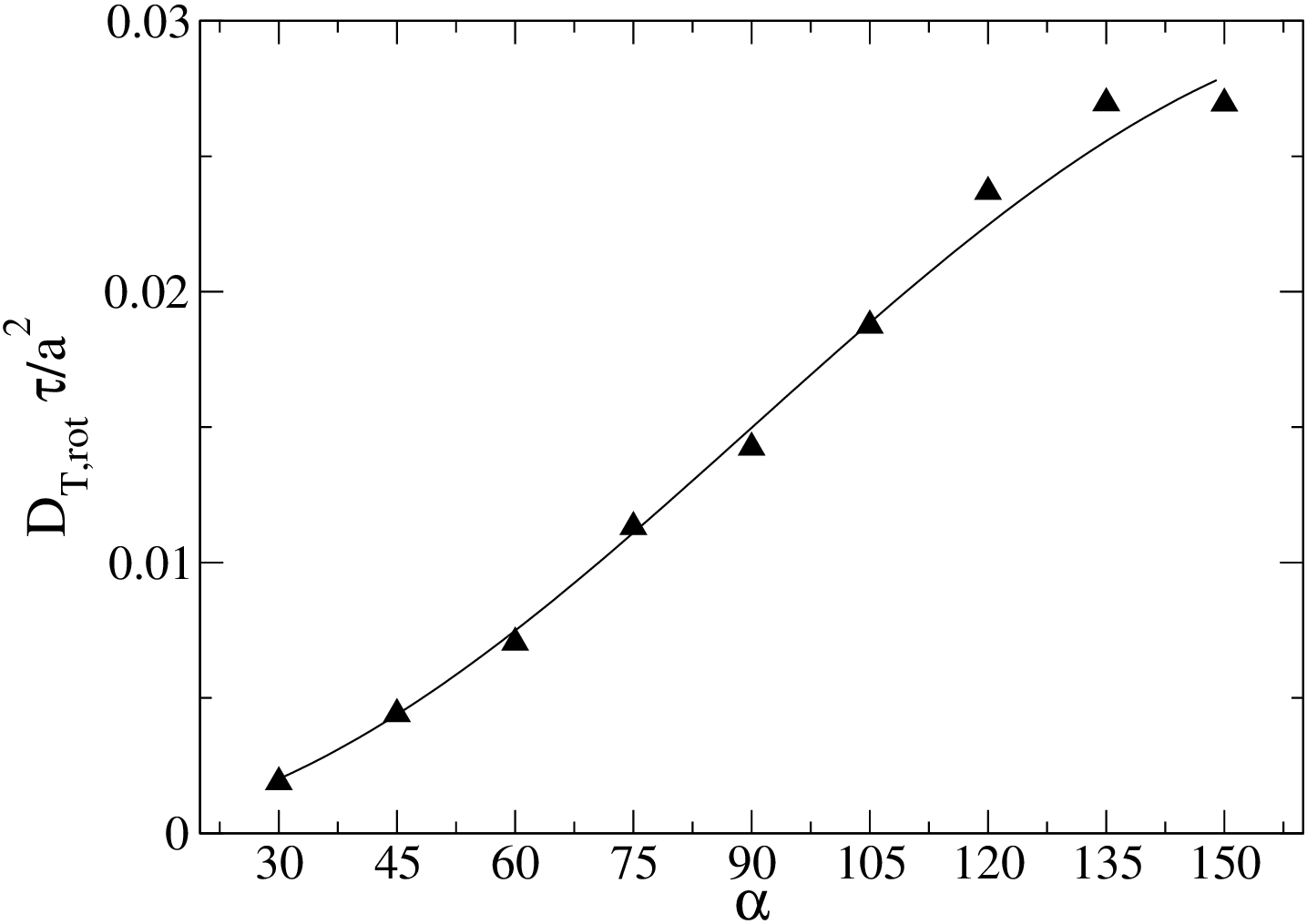}
\caption{The normalized collisional contribution to the thermal diffusivity, 
$D_T^{rot}\tau/a^2$, as a function of collision angle $\alpha$. 
The solid line is the theoretical prediction (\ref{DT_rot}).  
The data were obtained by time averaging over 360,000 iterations.
Parameters: $L/a=16$, $\lambda/a=0.1$, $M=3$ and $\tau=1$.}
\label{fig_dtrot}
\end{center}
\end{figure}

Using these results in (\ref{res2}) and averaging over the number of 
particles in a cell, assuming again that the probability of having a given 
particle, $i$, is in a cell containing a total number of particles 
$m_i$ is $m_iP_p(m_i,M)/M$, where $M$ is the average number of particles 
per cell, one finds  
\begin{eqnarray}
\label{DT_rot} 
D_T^{rot}&=&\frac{a^2}{3 (d+2) \tau} \frac{1}{M} 
\left[1-e^{-M}\left(1+\int_0^M \frac{e^x-1}{x}dx \right) \right] 
[1-\cos(\alpha)] \\
&\underset{\text{large $M$}}{\longrightarrow}&   
\frac{a^2}{3 (d+2) \tau} \frac{1}{M} 
\left(1 +{\rm e}^{-M}({\rm ln}M-1)- \frac{1}{M}-\frac{1}{M^2}-\frac{2}{M^3}-... \right) 
[1-\cos(\alpha)] \\
&\underset{\text{small $M$}}{\longrightarrow}&   
\frac{a^2}{12 (d+2) \tau}  
\left(M-\frac{5}{9}M^2+... \right) 
[1-\cos(\alpha)] 
\end{eqnarray} 
for all models considered. Note that in contrast to the viscosity, the 
rotational contribution to the thermal diffusivity is $O(1/M)$ for large $M$. 
Simulation results for the collisional contribution to thermal diffusivity
are compared with (\ref{DT_rot}) in Fig. \ref{fig_dtrot}. 
This contribution to the thermal diffusivity, which is not negligible for 
small M (such as $M=3$ in Fig. \ref{fig_dtrot}), was not discussed in 
Refs. \cite{kiku_03} and \cite{pool_05}. 

\begin{figure}
\begin{center}
\vspace{2cm}
\includegraphics[width=5in,angle=0]{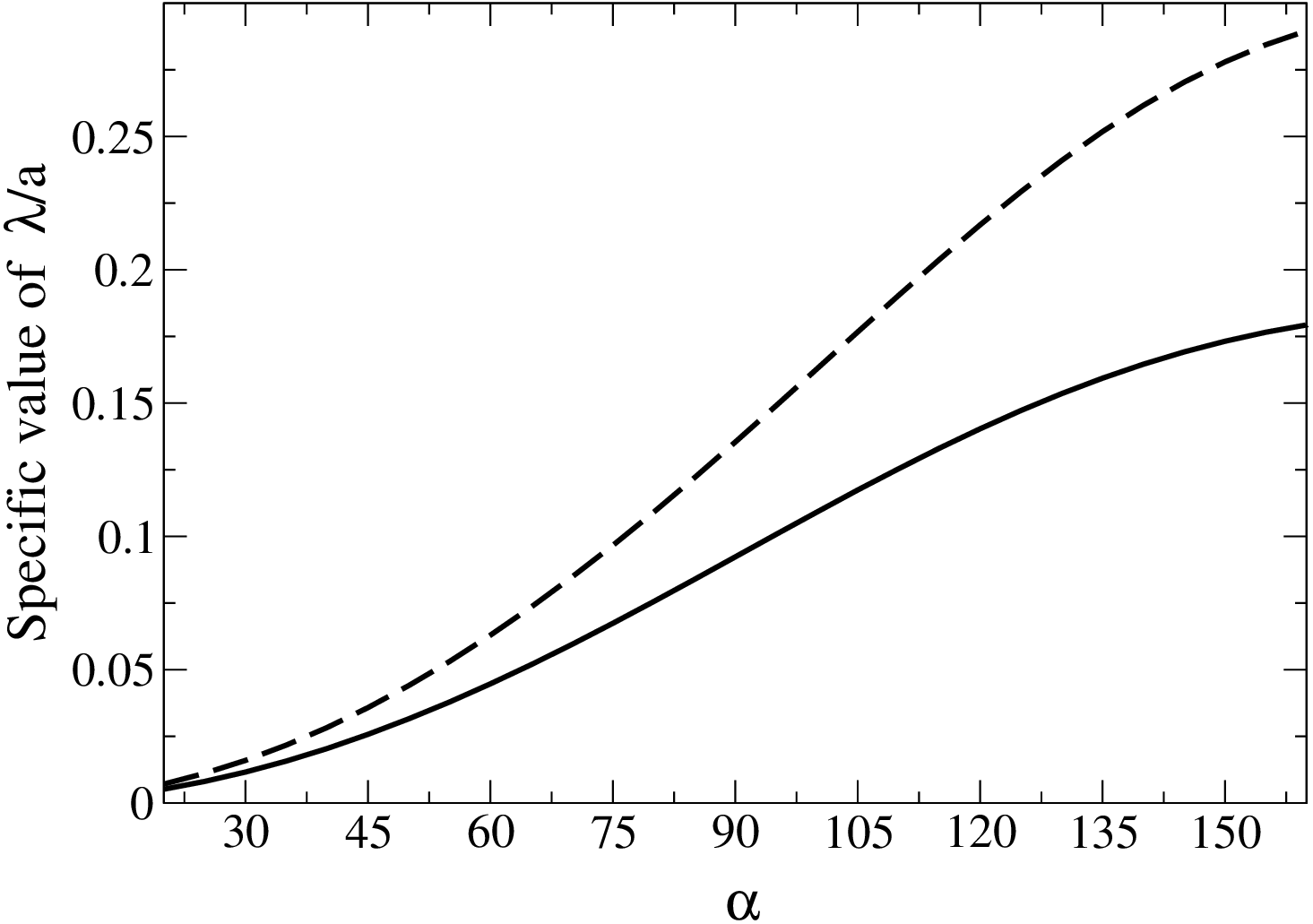}
\caption{The specific ratio, $\lambda/a$, at which 
$D_T^{kin}=D_T^{rot}$ as a function of the rotational angle $\alpha$ for $M=5$ 
in $d=2$ (dashed line) and $d=3$ (solid line). $\lambda=\tau\sqrt{k_B T}$ is the mean free path, $a$ is the cell size. 
For mean free paths below the 
curve the rotational contribution to the thermal diffusivity, $D_T^{rot}$,
is larger than the kinetic part, $D_T^{kin}$. Theoretical expressions given 
by Eq. (\ref{DT_rot}), Eq. (89) of Ref. \cite{ihle_03b}, and Eq. (48) of 
Ref. \cite{tuze_03} were used.
}
\label{fig_compar}
\end{center}
\end{figure}

A comparison of the relative size of the rotational and kinetic contributions 
to $D_T$ is given in Fig. \ref{fig_compar}. Since $D_T^{rot}$ is independent 
of temperature while $D_T^{kin}$ increases linearly with temperature, 
there is a temperature (or mean free path $\lambda=\tau\sqrt{k_B T}$) at which both contributions are 
equal. The ratio $\lambda/a$ of this specific mean free path to the cell size $a$ is plotted as a function 
of the rotation 
angle $\alpha$. Results obtained using Eq. (\ref{DT_rot}), Eq. (89) of 
\cite{ihle_03b}, and Eq. (48) in \cite{tuze_03} are presented for $M=5$ 
in both two (dashed line) and three dimensions (solid line). 
In two dimensions, $D_T^{rot}$ can be larger than $D_T^{kin}$ already at mean 
free paths as large as $0.25\,a$ for large $\alpha$. In $d=3$, the rotational 
contribution is slightly less important. 

\section{CONCLUSION}

It has been shown that the random shift procedure introduced in Refs. 
\cite{ihle_01,ihle_03a} not only restores Galilean invariance, but also 
enables an exact evaluation of the collisional contribution to the 
transport coefficients. The current approach justifies in detail several 
assumptions used in the non-equilibrium calculations of Ref. \cite{kiku_03} 
and \cite{pool_05}, and was used to determine the collisional contribution 
to the shear viscosity, the bulk viscosity, and the thermal diffusivity.

A detailed analysis of the consequences of the fact that SRD collisions  
do not conserve angular momentum was also presented. It was shown that while 
the long-time, long-length-scale hydrodynamics of the model are not affected, 
it does lead to small changes in the viscous contribution to the sound 
attenuation coefficient. Although it has been pointed out 
previously \cite{pool_05} that the collisional contribution to the macroscopic 
viscous stress tensor is not symmetric, our interpretation of the 
consequences of this fact is different from that of Ref. \cite{pool_05}. 
In particular, the resulting slight modification of the coefficient of 
sound attenuation has no consequences for most practical applications, 
such as those in Refs. \cite{falc_03b,nogu_04},  
and does not restrict the validity of the model. 

\section{Acknowledgement}
We thank J. Yeomans for helpful discussions which initiated this 
re-examination of the Green-Kubo approach. We also thank her and C.M. Pooley
for making their unpublished notes available to us, and Alexander Wagner 
for numerous discussions. Support from the 
National Science Foundation under grant Nos. DMR-0328468 and DMR-0513393, and 
ND EPSCoR through NSF grant EPS-0132289, 
is greatfully 
acknowledged. 

\section{Appendix} 

\subsection{$\langle (\Delta\xi^s_{ix})^2\rangle$} 

If $X_s=x_i+\delta$, with $0\le x_i < a$, 
\begin{equation} 
\langle \Delta\xi^s_{ix}\vert_X\rangle_\delta = \int_{-a/2}^{a/2} 
        [\Theta(-X_s)-\Theta(X_s-a)]d\delta
\end{equation} 
and 
\begin{equation} 
\langle (\Delta\xi^s_{ix})^2\vert_X\rangle_\delta = a\int_{-a/2}^{a/2} 
        [\Theta(-X_s)+\Theta(X_s-a)]d\delta .  
\end{equation} 
Integrating over $X$, we have 
\begin{equation} 
\label{DXI2}  
\langle (\Delta\xi^s_{ix})^2\rangle = \int_0^{a/2}d\delta  
    \int_{a-\delta}^a dx_i + \int_{-a/2}^0 d\delta \int_0^{-\delta} dx_i  
    =a^2/4. 
\end{equation} 

\subsection{$\langle \Delta\xi^s_{ix}\Delta\xi^s_{jx}\rangle$, $i\ne j$} 

\begin{equation} 
\langle \Delta\xi^s_{ix}\Delta\xi^s_{jx}\rangle = \frac{1}{a}
\int_{-a/2}^{a/2}d\delta\left\{\int_0^a dx_i[\Theta(-x_i-\delta)-
                 \Theta(x_i+\delta-a)]\right\}^2. 
\end{equation} 
The integral over $x_i$ is 
\begin{equation} 
\int_0^a dx_i[\Theta(-x_i-\delta)-\Theta(x_i+\delta-a)] = 
\delta \Theta(\delta)-\delta\Theta(-\delta), 
\end{equation} 
so that 
\begin{equation} 
\label{DXIJ} 
\langle \Delta\xi^s_{ix}\Delta\xi^s_{jx}\rangle = 
\frac{1}{a}\int_{-a/2}^{a/2}d\delta[\delta^2\Theta(\delta)+ 
\delta^2\Theta(-\delta)] = a^2/12.
\end{equation} 

\subsection{Various other correlations}

There are a number of other useful relations which are required to evaluate 
the $\bf{B}$-correlations which can be easily evaluated using the same techniques. 
They include 
\begin{equation} 
\label{a1} 
\langle\Delta\xi_{ix}^s v_{ix}\rangle = 
\langle\Delta\xi_{ix}^s(-\tau) v_{ix}\rangle = 0   
\end{equation} 
and 
\begin{equation}
\label{a2} 
\langle\Delta\xi_{ix}\Delta\xi_{jx}\rangle = 
\langle\Delta\xi_{ix}\Delta\xi_{jx}^s\rangle = 0,   
\end{equation} 
\begin{equation} 
\label{a3} 
\langle\Delta\xi_{ix}^s v_{jx}\rangle = 
\langle\Delta\xi_{ix}^s(-\tau) v_{jx}\rangle = 0,  
\end{equation}  
and 
\begin{equation} 
\label{a4} 
\langle\Delta\xi_{ix}^s\Delta\xi_{jx}^s(-\tau)\rangle = 
\langle\Delta\xi_{ix}\Delta\xi_{jx}^s(-\tau)\rangle = 0     
\end{equation}  
for $i\ne j$. 
Finally, 
\begin{equation} 
\label{a5} 
\langle\Delta\xi_{ix}^s(-\tau)[\Delta\xi_{ix}(0)-\Delta\xi_{ix}^s(0)]\rangle 
=-a^2/12.
\end{equation} 
To show this, assume $0\le x_i(0)<a$ and $na\le x_i(0)+\tau v_{ix}(0)<(n+1)a$. 
Averaging first over $\delta$ at time $\tau$ and then the random shift 
at time $t=0$, for fixed $x_i(0)$ and $v_{ix}(0)$, one finds 
\begin{equation}
\label{p1}  
\langle\Delta\xi_{ix}^s(-\tau)[\Delta\xi_{ix}(0)-\Delta\xi_{ix}^s(0)]
\vert_{x_i(0),X}\rangle = \langle(a/2-x_i(0))(-a/2+x_i(0)-\tau v_{ix}(0)\rangle.
\end{equation} 
The final ensemble average in (\ref{p1}) reduces to  
\begin{eqnarray} 
\nonumber
& &\frac{1}{a}\int_0^a dx \sum_{n=-\infty}^\infty (a/2-x)  
\int_{(na-x)/\tau}^{[(n+1)a-x]/\tau} dv_x [-a/2+x+\tau v_x]
w(v_x) \\
 &=& \frac{1}{a}\int_0^a dx\, (a/2-x) \int_{-\infty}^\infty dv_x  
[-a/2+x+\tau v_x]w(v_x) ,    
\end{eqnarray} 
where we have dropped the index $i$ and the time argument of $x$ and $v_x$ 
for brevity. The integral over $v_x$ can be performed immediately, and 
the remaining integral over $x$ gives the result, $-a^2/12$. 

\subsection{Proof of relation $\langle v_{jx}(\tau)v_{jy}(\tau)  
v_{ix}(0)B_{iy}(0)\rangle=0$.} 

In order to evaluate this expression, averages over the random shift at 
time $t=0$, $\delta_0\equiv(\delta_{0x},\delta_{0y})$, the shift at time 
$t=\tau$, $\delta_\tau\equiv(\delta_{\tau x},\delta_{\tau y})$, and over 
the initial positions, ${\bf r}_i(0)=(x_i(0),y_i(0))$, and velocities, 
${\bf v}_i(0)$ of {\it all} particles are required. 
Averaging first over $\delta_0$, keeping all the other quantities fixed,  
yields Eq. (\ref{ex1}). Next, note that $\xi_{iy}^s(\tau)$ has an implicit 
dependence on the initial positions and velocities at $t=0$ and 
$\delta_\tau$. We therefore write 
\begin{equation} 
\xi_{iy}^s(\tau)=
\xi_{iy}^s(\tau;\delta_\tau,\{{\bf r}_k \},\{{\bf v}_k \}).  
\end{equation} 

Because of translational symmetry, 
\begin{equation} 
\xi_{iy}^s(\tau;\delta_\tau,\{{\bf r}_k \},\{{\bf v}_k \}) 
\equiv \xi_{iy}^s(\tau;0,\{{\bf \tilde{r}}_k \},\{{\bf v}_k \}),  
\end{equation} 
with ${\bf \tilde{r}}_k={\bf r}_k+\delta_\tau$.
Keeping $\{{\bf \tilde{r}}_k\}$ fixed,  
the average over $\delta_\tau$ in (\ref{ex1}) then become 
\begin{eqnarray}
\nonumber
& & \frac{1}{a}\prod_{k=1}^N \int_{-\infty}^\infty f(\{{\bf v}_k\})\, 
d{\bf v}_k\int_{-\infty}^\infty\, d{\bf r}_k\,\int_{-a/2}^{a/2}\,
v_{jx}(\tau)v_{jy}(\tau)v_{ix}(0)
\left[\xi_{iy}^s(\tau)-y_{i}(0)+a/2-\tau v_{iy}(0)\right]\,d\delta_\tau \\
\nonumber &=&
\frac{1}{a}\prod_{k=1}^N \int_{-\infty}^\infty f(\{{\bf v}_k\})\,d{\bf v}_k\, 
\int_{-\infty}^\infty\,d{\bf \tilde{r}}_k\,\int_{-a/2}^{a/2}\,
v_{jx}(\tau)v_{jy}(\tau)v_{ix}(0)\left[\xi_{iy}^s(\tau)-
\tilde{y}_{i}(0)+\delta_\tau+a/2-\tau v_{iy}(0)\right]\,d\delta_\tau \\ &=&
\prod_{k=1}^N \int_{-\infty}^\infty f(\{{\bf v}_k\})\,d{\bf v}_k\, 
\int_{-\infty}^\infty\,d{\bf \tilde{r}}_k\,
v_{jx}(\tau)v_{jy}(\tau)v_{ix}(0)\left[\xi_{iy}^s(\tau)-\tilde{y}_{i}(0)
+a/2-\tau v_{iy}(0)\right]\,.  
\end{eqnarray}
where $f$ is the N-particle Boltzmann distribution.

The remaining average over the initial configuration can be split up into a 
sum of several terms.  Each term corresponds to a situation in which the 
particle 
labelled $i$ is restricted at time zero to be in a specific cell $\xi_1$ 
together with $k_1$ other particles with given labels, while particle
$j$ is likewise residing only in a given cell $\xi_2$ together with a set 
of $k_2$ distinguishable particles. These restrictions are needed in order 
to have the post-collisional velocities ${\bf v}_j(\tau)$ and the cell label 
$\xi_{jy}^s(\tau)$ to be independent on the initial positions of the particles 
in every term.

We will show now that all these terms will vanish independently.
Keeping  the initial velocities fixed, an average over $\tilde{y}_i(0)$ is 
performed under the condition mentioned above, i.e. where 
$\xi_{jy}^s(\tau)=\xi_1$.  These constraints mean that $\tilde{y}_i(0)$ 
must be in the interval 
$[\xi_1-\tau v_{iy}(0), \xi_1+a-\tau v_{iy}(0)]$.
Since the post-collisional velocities of particle $j$ are not affected by 
the position $\tilde{y}_i(0)$, each of the terms are proportional to
\begin{equation}
\int\, f(\{{\bf v}_i\})\, d{\bf v}_i^N\,\int_{\xi_1-v_{iy}(0)\tau}^
{\xi_1+a-v_{iy}(0)\tau}\;  
\left[ \xi_1+\frac{a}{2}-\tilde{y}_i(0)-\tau v_{iy}(0)\right]\,d\tilde{y}_i(0)
\propto \left.a[\xi_1+\frac{a}{2}-\tau v_{iy}(0)]-\frac{\tilde{y}_i(0)^2}{2}
\right|_{\xi_1-\tau v_{iy}(0)}^{\xi_1+a-\tau v_{iy}(0)}=0\,.  
\end{equation}
All terms vanish independently, so that 
$\langle v_{jx}(\tau)v_{jy}(\tau) v_{ix}(0)B_{iy}(0)\rangle$ 
is zero. Note that the same argument applies when $\xi_2=\xi_1$, i.e. the 
particles being in the same cell, for both $j\ne i$ and $j=i$.

\subsection{Entropy production}

The {\it general equation of heat transfer} is \cite{land_59}

\begin{equation}
\label{ENTRO1}
\rho T\left(\frac{\partial s}{\partial t}+{\bf v \cdot \nabla}\,s\right)
=\hat{\sigma}_{\alpha\beta}\,\partial_{\beta}v_\alpha+
\nabla(\kappa{\bf \nabla} T)\,,  
\end{equation}
where the stress tensor $\hat{\sigma}_{\alpha\beta}$ is given in Eq. 
(\ref{stress}), $s$ is the entropy per unit mass, $\rho$ is the mass density, 
and $\kappa$ is the thermal conductivity. It is easily verified that
$\hat{\sigma}_{\alpha\beta}\,\partial_{\beta}v_\alpha$ can be written as  
\begin{equation}
\label{ENTRO2}
\hat{\sigma}_{\alpha\beta}\,\partial_{\beta}v_\alpha=\frac{\nu_1}{2}
\left(\partial_\beta v_\alpha+\partial_\alpha v_\beta-
\frac{2}{d}\delta_{\alpha\beta}\partial_\gamma v_\gamma \right)^2+
\frac{\nu_2}{2}(\nabla\times{\bf v})^2 +\gamma(\nabla\cdot {\bf v})^2\,, 
\end{equation}
where all three terms are independent, invariant under coordinate 
transformations, and can vanish independently.
Using (\ref{ENTRO1}) and (\ref{ENTRO2}), it can be shown \cite{land_59} that 
the entropy production in a given volume $\Omega$ is
\begin{equation}
\label{ENTRO3}
\frac{d}{dt}\,\int_\Omega\,\rho s\,dV=\int_\Omega\left[
\frac{\kappa (\nabla T)^2}{T^2} + \frac{\nu_1}{2T}
\left(\partial_\beta v_\alpha+\partial_\alpha v_\beta-\frac{2}{d}
\delta_{\alpha\beta}\partial_\gamma v_\gamma \right)^2
+\frac{\nu_2}{2T} (\nabla\times{\bf v})^2 + \frac{\gamma}{T} 
(\nabla\cdot{\bf v})^2 \right]\,dV\,.   
\end{equation}
The requirement that the entropy production is non-negative implies that 
all transport coefficients in (\ref{ENTRO3}) are greater than or equal to zero, 
namely, 
\begin{equation}\label{FR}  
\nu_1\geq0,\ \ \ \ \nu_2\geq0, \ \ \ \ \gamma\geq0,\ \ \  {\rm and} \ \ \ \ 
\kappa\geq0.
\end{equation} 
For $\nu_2=0$ this condition is identical to the well-known result 
discussed in \cite{land_59}. For SRD, conditions (\ref{FR}) reduce to 
\begin{equation}      
\nu^{kin}+\nu_1^{rot}\geq0,\ \ \ \ \nu_2^{rot}\geq0, \ \ \ {\rm and}  
\ \ \ \ \gamma^{rot}\geq0.   
\end{equation}

\newpage

\newpage

\vspace{10cm}
\newpage 

\newpage

\newpage

\begin{thebibliography}{99}
\bibitem{pol_fl} There are, in fact, several types of fluids in which the 
angular momentum is not conserved. Examples include polar fluids, poly-atomic 
fluids subject to body torques, and certain non-Newtonian fluids~\cite{aris_89}.
\end{thebibliography}
\end{document}